\documentclass[journal]{IEEEtran}
\usepackage[pdftex]{graphicx}
\usepackage{amsmath}
\usepackage{amsthm}
\usepackage{latexsym}
\usepackage{multirow}

\begin{document}

\title{A New Noncoherent Decoder for Wireless Network Coding}

\author{Sunjoo Moon and Cong Ling
\thanks{S. Moon and C. Ling are with the Department
of Electrical and Electronic Engineering, Imperial College London, London, SW7 2AZ, UK (e-mail:
sunjoo.moon08@imperial.ac.uk, cling@ieee.org).} }
\maketitle%
\markboth{}{}%

\begin{abstract}
This work deals with the decoding aspect of wireless network coding in the canonical two-way relay
channel where two senders exchange messages via a common relay and they receive the mixture of two
messages. One of the recent works on wireless network coding was well explained by Katti
\textit{et al.} in SIGCOMM'07. In this work, we analyze the issue with one of their decoders when
minimum-shift keying (MSK) is employed as the modulation format, and propose a new noncoherent
decoder in the presence of two interfering signals.\newline

\emph{Index terms}: network coding, wireless networks, cooperative transmission, relay network.
\end{abstract}

\section{Introduction}

Before the benefits of coding for networking were recognized, wireless networks were designed to
protect packets from interference. Interference was considered a factor which has only negative
effects on network systems. However, network coding \cite{2} in the wireless context, termed
analog network coding (ANC) \cite{1} or physical-layer network coding (PNC) \cite{9}, breaks this
concept. Rather, intended interference is encouraged at the relays for high throughput. The main
idea of network coding is that information, not packets, is the factor to be sent to a
destination. It means that although packets may collide, there is no problem if the information,
which each destination wants to receive, can be obtained. For this reason, network throughput can
be increased by sending collided messages instead of transmitting messages separately.

This work deals with ANC introduced by Katti \textit{et al.} \cite{1}. In ANC, analog signals,
which are generally represented as complex numbers, are mixed at the router instead of bits. By
combining two analog signals simultaneously at the relay located in the middle of two senders, the
interfered signal is forwarded to two transmitters. Two noncoherent decoders were proposed in
\cite{1}. In one of them the amplitude may be estimated directly from a signal clear of
interference: we refer to it as the direct method. The other decoder estimates amplitudes from
interfered signals: we refer to it as the joint method. The former decoder is noisier than the
latter. However, we believe the joint method was not used when it came to implementation; rather,
they obtained amplitudes from the free interference part \cite{20}.

In this paper, we first analyze the issue of the joint method. It turns out that the joint method
has difficulty in estimating the amplitudes when minimum-shift keying (MSK) was employed as the
modulation format. We then propose a new method of joint estimation of the signal amplitudes.
Other parts of our decoder are the same as \cite{1}, and our decoder has almost the same error
performance as the direct method.


Several improved algorithms avoiding coherent demodulation have been introduced. A decoder for
frequency-shift keying (FSK) systems is introduced in \cite{19} which requires a larger
signal-to-noise ration (SNR) for the same throughput compared with the counterpart. The
noncoherent decoders in \cite{17,18} assumed the knowledge of average received energies or the
statistics of channel attenuation. This work, however, makes no such assumptions; instead, it
obtains received energy from the interference signal.

\section{Illustrating example and system model}

\begin{figure}[t]
\centering
\includegraphics[scale=0.4]{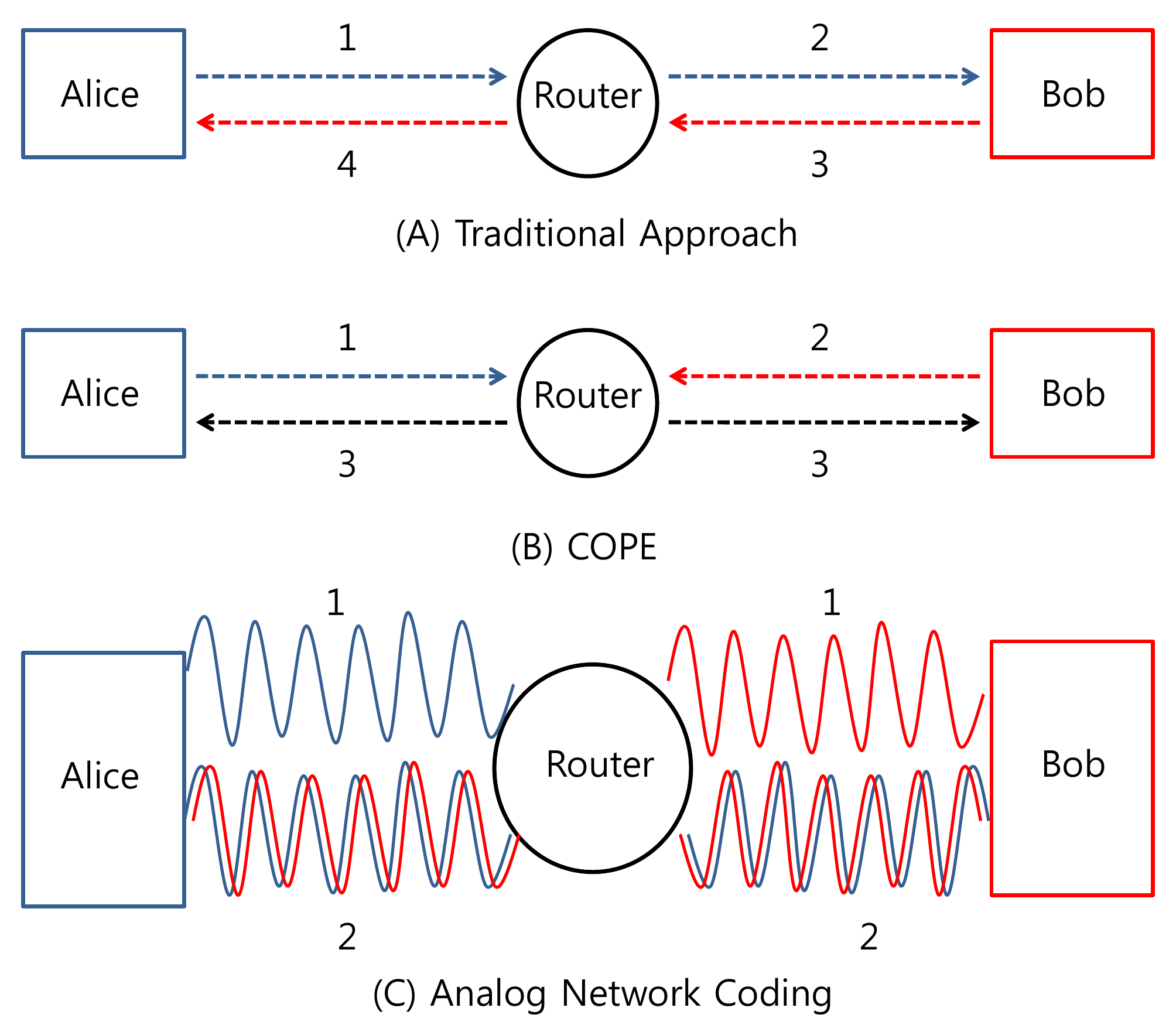}  \caption{Alice-Bob topology: (a) traditional approach, (b) digital network coding (COPE \cite{8}), (c), ANC. With ANC, the capacity of the Alice-Bob topology can be improved twice as much as that of traditional approach.}\label{fig:1}
\end{figure}

Consider the canonical 2-way relay network which is well known as Alice-Bob topology and only one
packet is allowed to be sent at once. When Alice and Bob want to exchange a pair of packets
through a relay as Fig. \ref{fig:1} shows, total 4 time slots are used for the traditional
approach. Alice sends a message to the router and the router forwards it to Bob. The process of
Bob is also similar to Alice's process. By digital network coding \cite{8}, total time slots can
be reduced to 3. Alice sends a packet and Bob sends a packet sequentially then the router XORs the
two packets and forwards it to both destinations. By XORing with their own packet, Alice and Bob
can obtain the packet which they want to receive. Interestingly, total time slot can be reduced to
2 by ANC \cite{1}. Two senders transmit the packets simultaneously. If Alice and Bob can decode
what they want to receive from the collided packet, the throughput can be increased.

ANC proposed in \cite{1} uses the MSK modulation. In order to understand how ANC works, a brief
instruction about the modulation and the form of the signal at both sender and receiver side will
be given.

\subsection{The transmitter side}
In the baseband equivalent model, a wireless signal is represented as a complex number which has
a phase and an amplitude and is a discrete-time function. In this sense, the transmitted signal
from the sender takes the form

\begin{center}
    $S[n]=A_s[n]e^{i \theta_s[n]}$
\end{center}
where the subscript $s$ means sender side. We need to map ``0" and ``1" to two different forms of
complex functions.

The MSK modulation uses a phase difference between consecutive complex samples to distinguish bits. When the next bit is ``1" then gives a $\frac{\pi}{2}$ phase difference comparing with the previous one while a phase difference of $-\frac{\pi}{2}$ represents ``0".

\subsection{The receiver side}
What we are interested in is how the received signal looks like under the condition which the
channel is quasi-static flat-fading. An amplitude and a phase in the transmitted signal are
different from those of the received signal due to channel attenuation and a phase shift. When the
transmitted signal is $A_s[n]e^{i\theta_s[n]}$, we can represent the received signal as

\begin{center}
    $y[n]=hA_s[n]e^{i(\theta_s[n]+\gamma)}$
\end{center}
where $h$ is channel attenuation and $\gamma$ is a phase shift which varies according to the distance between the sender and the receiver.

Although there is a fading, the decoding for the MSK is simple. When we map the received complex
samples back into the bit stream in MSK, the phase difference between consecutive complex samples
decides whether the bit is ``0" or ``1". The amplitude for the MSK is constant so consecutive
complex samples look like $hA_se^{i(\theta_s [n]+\gamma)}$ and $hA_s e^{i(\theta_s
[n+1]+\gamma)}$. In the noiseless case, simply we can know the phase difference between
consecutive samples by

\begin{center}
    $r=\frac{hA_s e^{i(\theta_s [n+1]+\gamma)}}{hA_s e^{i(\theta_s [n]+\gamma)}}=e^{i(\theta_s[n+1]-\theta_s[n])}.$
\end{center}

For that reason, if the angle of $r$ is $\frac{\pi}{2}$ then we map it to ``1" and
$-\frac{\pi}{2}$ to ``0". However, in practice, the angle of complex number $r$ is not exactly
$\frac{\pi}{2}$ or $-\frac{\pi}{2}$ since there are channel noise and estimation errors.
Therefore, if the phase difference is greater than 0 then it is mapped to ``1" and if the phase
difference is negative then it denotes ``0".

\section{Analysis of Existing Decoder}
The received signal at a destination is the sum of Alice and Bob's signals after amplifying by the
router. However, the interfered signal is not the exactly same as the sum of two signals which
Alice and Bob sent. The reason is that two transmitted signals experience the channel from each
sender to the router and the channel from the router to the corresponding receiver. As a result,
the signal that they receive is \cite{1}:
\begin{eqnarray}
  \nonumber y[n]  &=& y_A [n]+ y_B [n] \\
  \nonumber &=& h' A_s e^{i(\theta_s [n]+\gamma')} +h''B_se^{i(\phi_s [n]+\gamma'' )}
\end{eqnarray}
where $\theta_s$ and $\phi_s$ refer to the phase of the signal transmitted by Alice and Bob,
respectively, whereas $A_s$ and $B_s$ are the amplitudes.


The simple decoding is that Alice estimates the channel parameters $h'$, and $\gamma'$ then
creates the version of her own signal and subtract it from the received signal. The remainder is
$y_B[n]$ which Alice can decode by the standard MSK demodulation. However, this technique suffers
a few drawbacks. It incurs overhead, especially when the channel is time varying; also, estimating
phase shift is not simple and requires accurate coherent phase tracking.


In \cite{1}, there is a two-step process introduced to calculate phase differences when two signals interfere. In order to make it simple, let the representation of the received signal be :
\begin{equation}
y[n]= Ae^{(i\theta[n])}+Be^{(i\phi[n])}
\label{eq:1}\end{equation}
where $A=h'A_s, B=h''B_s,\theta[n]=\theta_s[n]+\gamma'$, and $\phi[n]=\phi_s[n]+\gamma''$.

\subsection{Possible phases}
 It is not straightforward for Alice to tell the exact values of $\theta[n]$ and $\phi[n]$ just by
 analyzing the interfered signal. However, once she knows the amplitudes, $A$ and $B$, then she can calculate
 possible values of $\theta[n]$ and $\phi[n]$ by the following lemma proved in \cite{10,11}.\newline

LEMMA 1. If $y[n]$ is a complex number satisfying (\ref{eq:1}), then the pair $(\theta[n],
\phi[n])$ takes one of the following two pairs \cite{1}.
\begin{equation}
\theta[n]=\arg(y[n](A+BD \pm iB\sqrt{1-D^2})) \label{eq:2}\end{equation}
\begin{equation}
\phi[n]=\arg(y[n](B+AD \mp iA\sqrt{1-D^2})) \label{eq:3}\end{equation} where $D=
\frac{(|y[n]|^2-A^2-B^2)}{2AB}$, and $\arg$ is the angle of the complex number.

For each solution to $\theta[n]$, there is a unique solution for $\phi[n]$. For instance, if
$\theta[n]=\arg(y[n](A+BD+iB\sqrt{1-D^2}))$, then $\phi[n]=\arg(y[n](B+AD-iA\sqrt{1-D^2}))$.
Accordingly, the lemma gives two pair of solutions.


\subsection{Existing method to estimate amplitude A and B}
In order to obtain possible phase pairs using the Lemma 1, two equations were introduced in
\cite{1}.

The energy of the interfered signal is
\begin{equation}\label{energy}
    E[|y[n]|^2]=E[A^2+ B^2+ 2AB\cos(\theta[n]-\phi[n] )]
\end{equation}
where $E[\cdot]$ is the expectation. If the input bits are random, $E[2AB\cos(\theta[n]-\phi[n])]$
will be 0. For this reason, the input bits were XORed with a pseudo-random sequence before the
modulation and then XORed again at the receiver to get original bits in \cite{1}. Therefore, the
first equation can be reduced to
\begin{equation}\label{eq:4}
\mu = E[|y[n]|^2 ]= A^2+ B^2.
\end{equation}

Alice can estimate $\mu $ by observing the average energy in the received signal. The second
equation comes from the variance of the received signal. Alice can calculate the following
quantity \cite{1}:
\begin{equation}\label{eq:sigma}
    \sigma=  \frac{2}{N} \sum_{|y[n]|^2>\mu}|y[n]|^2.
\end{equation}
The value for $\sigma$ was shown to be \cite{1}
\begin{equation}
\sigma=A^2+B^2+\frac{4AB}{\pi}. \label{eq:5}
\end{equation}
From the two equations, Alice can
estimate two amplitudes.

However, we realize that, \emph{in the case of MSK modulation}, the value for $\sigma$ cannot be
reduced to (\ref{eq:5}). The reason is as follows.

The value $\sigma$ means the average energy calculated from the received complex samples that is
greater than mean value. By (\ref{energy}), we can rewrite $\sigma$ as

\begin{center}
    $\sigma=E[|y[n]|^2 |\cos(\theta[n]-\phi[n] )>0]$


    $=A^2+ B^2+ 2ABE[\cos(\theta[n]-\phi[n])|\cos(\theta[n]-\phi[n]  )>0].$
\end{center}

If one took the average of a cosine over its positive lobes,
$E[\cos(\theta[n]-\phi[n])|\cos(\theta[n]-\phi[n] )>0]$ would be reduced as $\frac{2}{\pi}$. But
this is not true in the case of MSK modulation because the phases can only take \emph{discrete
values}.

Let $R$ be the initial angle between two vectors. There are 4 possible pairs for the next
interfered sample. If the next bit for Alice and Bob is the same like 1 and 1 or 0 and 0, the
phase for the two vectors will shift to the same way, $\frac{\pi}{2}$ or $-\frac{\pi}{2}$. It
means that $R$ does not change when the next bits for two signals are the same. If Alice's next
bit is 1 and Bob's next bit is 0 then $R$ will change to $\pi+R$. when Alice's next bit is 0 and
Bob's next bit is 1 then $R$ will change to $\pi-R$. That is what we can make all the possible
angles in the next interfered sample. However the values for $\cos(\pi+R)$ and $\cos(\pi-R)$ are
the same (i.e., $-\cos(R)$). Hence, there are only two possibilities for the value of a cosine,
$\cos(R)$ or $-\cos(R)$. Therefore, if $\cos(R)$ is positive, $E[\cos(\theta[n]-\phi[n]
)|\cos(\theta[n]-\phi[n] )>0]=\cos(R)$; if $\cos(R)$ is negative,
$E[\cos(\theta[n]-\phi[n])|\cos(\theta[n]-\phi[n] )>0]=-\cos(R)$. This pattern is repeated during
the same packet. That is, there is a fixed value
\begin{equation}\label{}
    E[\cos(\theta[n]-\phi[n] )|\cos(\theta[n]-\phi[n] )>0] = |\cos(R)|
\end{equation}
and therefore
\begin{equation}
\sigma=A^2+B^2+2AB|\cos(R)| \label{eq:truesigma}
\end{equation}
throughout the same packet and it will not be changed until the next packet. Therefore, in general
(\ref{eq:5}) does not hold within a given packet for MSK modulation\footnote{One might wonder if
(\ref{eq:5}) could hold true by averaging over many packets (such that the angle $R$ will be
truely random on $[0,2\pi]$). However, this does not work since the amplitudes will also change
from packet to packet due to fading.}.

Moreover, even if we assume that (\ref{eq:5}) were right, there is still an ambiguity which
amplitude is for Alice. This is because (\ref{eq:4}) and (\ref{eq:5}) are completely symmetric
with respect to $A$ and $B$.
%
%
%
Without extra information we cannot distinguish which amplitude is for Alice. If we apply the
wrong amplitude to Alice, it will cause a problem.

Consider the equations (\ref{eq:2}) and (\ref{eq:3}), which are used to obtain possible phases. If
we exchange $A$ and $B$, the equation for $\theta$ is changed to the equation for $\phi$. That is,
if we apply wrong amplitude to Alice, it will cause a problem where Alice would align her known
phase difference with the phase difference for Bob. Therefore, Alice has to distinguish the
amplitude for her signal.

For these reasons, we need a new method to estimate the amplitudes\footnote{When it was presented
at the SIGCOMM'07 conference, the joint method was not mentioned \cite{20}.}.

\subsection{Estimating phase differences}

Once Alice get two possible pairs of phases, she needs to pick right pair of phases. In \cite{1}, the way to choose right
phase difference was well explained. When the two solution pairs are represented as
$(\theta_1[n],\phi_1 [n])$ and $(\theta_2 [n],\phi_2 [n])$, Alice has the following four possible
phase difference pairs \cite{1}.
\begin{flushleft}
   $(\Delta\theta_{xy}[n],\Delta\phi_{xy}[n])= $
\end{flushleft}
\begin{equation}
(\theta_x[n+1]-\theta_y[n],\phi_x [n+1]-\phi_y [n] ),\;\;\;\;\;\;\forall x,y \in \{1,2\}.
\label{eq:8}\end{equation}

Alice knows the phase difference of her signal and thus she uses it to pick up the right phase
difference pair. Alice compares the phase difference of her signal, $\theta_s [n]$ with the four
possible phase differences using the next following equation to pick up the correct
$\theta_{xy}[n]$ \cite{1}:
\begin{equation}
\text{err}_{xy}=|\Delta\theta_{xy}[n]-\Delta\theta_s[n]|,\;\;\;\forall x,y\in \{1,2\}.
\label{eq:9}\end{equation}

Alice can determine the correct $\theta_{xy} [n]$ by choosing the smallest error from (\ref{eq:9})
then she finds the unique $\phi_{xy} [n]$.

\section{The new method to estimate amplitudes}



The key point for estimating amplitudes from the interfered signal is to find the factor which
changes according to the amplitudes. The answer lies in transformation of the structure in the
interfered signal when angle between two signals changes from the constructive form to the
destructive form or vice versa. The transformation occurs when the next
bits for Alice and Bob' signals are different. The transformation between two interfered signals
$(X_1, X_2)$ contains the information about two amplitudes.

%
%
%

\begin{figure}[t]
\centering
\includegraphics[scale=0.3]{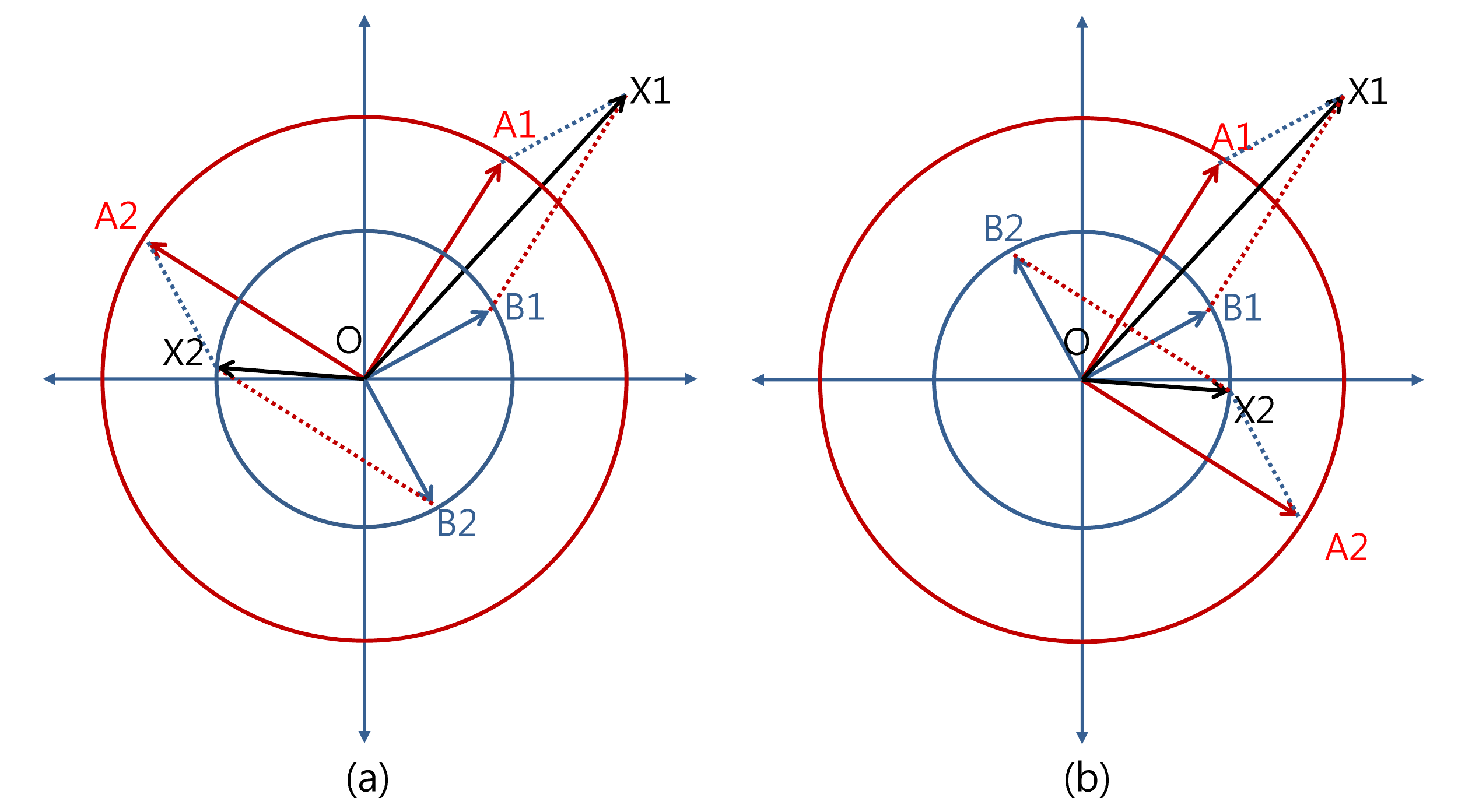}  \caption{The examples of the forms of two complex samples and possible vectors when their forms are changed from constructive to destructive.}\label{fig:7}
\end{figure}

\begin{figure}
\centering
\includegraphics[scale=0.3]{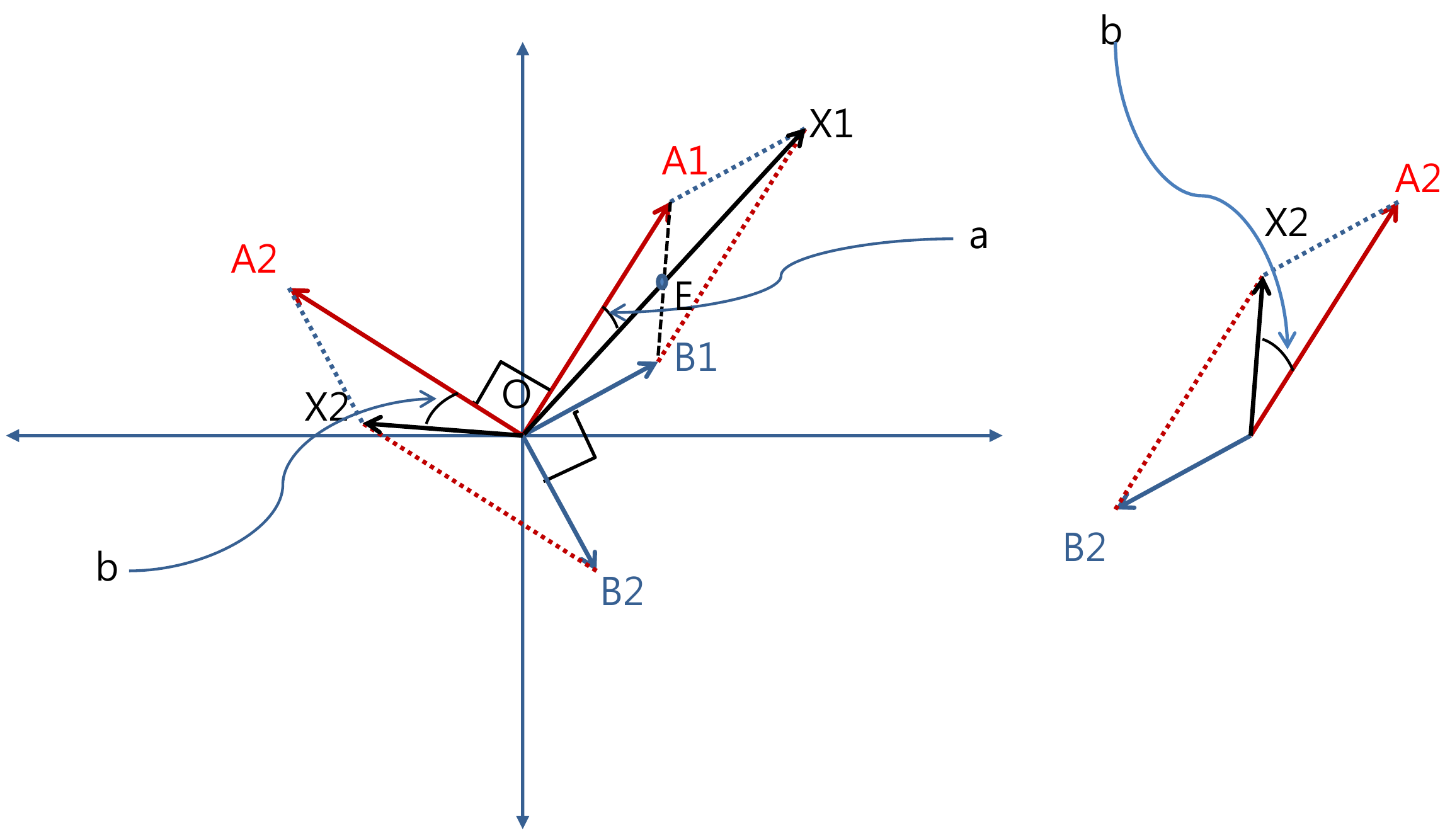}  \caption{Two parallelograms in consecutive bits which changes the form.}\label{fig:8}
\end{figure}

Consider Fig. \ref{fig:7}. $A_1$ denotes the first vector for Alice and $A_2$ the second while
$B_1$ and $B_2$ mean Bob's vectors. Assume that the next bit for Alice is 1 and the next bit for
Bob is 0 so that the form changes from constructive to destructive. When we look at the two groups
of vectors in detail, we can find the two parallelograms which are $\Diamond {O A_1 X_1 B_1}$ and
$\Diamond {O A_2 X_2 B_2}$. In Fig. \ref{fig:8}, the two parallelograms are described in more
details; so from now we will explain the new method based on Fig. \ref{fig:8}. If we know
$\overline{OE}, \overline{EA}$ and $\angle {A_1 E O}$, then we can calculate the length of $A_1$
which is the amplitude for Alice signal.

As Fig. \ref{fig:8} shows, $\angle{A_2 O X_2}=\angle{A_1 B_1 X_1}$ and $\angle {A_1 B_1
X_1}=\angle {O A_1 B_1}$. From the received signal, we can get the angle difference between $X_1$
and $X_2$ which is $\angle {X_2 O X_1}$. It means that we can get $\angle{A_1 E O}$ by $\pi -
(\angle{X_2 O X_1} - \pi/2)$.

Since the diagonals of a parallelogram bisect each other, $\overline{OE}=
\frac{\overline{OX_1}}{2}$. In the same way $\overline{EA_1}=\frac{\overline{OX_2}}{2}$. Now the
amplitude of Alice signal is
$\sqrt{\overline{OE}^2+\overline{EA_1}^2-2(\overline{OE})(\overline{EA_1})\cos(\angle{ A_1 E
O})}$. In addition, if we represent it just using the information of the received complex samples
then it can be rewritten as



\begin{equation}\label{}
    A = \sqrt{(\frac{\overline{OX_1}}{2})^2+(\frac{\overline{OX_2}}{2})^2
-2(\frac{\overline{OX_1}}{2})(\frac{\overline{OX_2}}{2})\cos({\pi - (\angle {X2 O X_1} -
\pi/2)})}\nonumber
\end{equation}

\begin{equation}
= \sqrt{\frac{1}{4}(X_1^2+X_2^2)+\frac{1}{2}(X_1X_2)\sin({\angle {X_2 O X_1} })}.
\label{eq:6}\end{equation}

In the same way, the amplitude for Bob's signal is
\begin{equation}
B=\sqrt{\frac{1}{4}(X_1^2+X_2^2)-\frac{1}{2}(X_1X_2)\sin({\angle {X_2 O X_1}})}.
\label{eq:7}\end{equation}

The derivation is similar in other cases (e.g., the next bit for Alice is 0 and the next bit for
Bob is 1). To summarize, $A$ and $B$ take values in one of the following two pairs
\begin{eqnarray}
  A &=& \sqrt{\frac{1}{4}(X_1^2+X_2^2)\pm \frac{1}{2}(X_1X_2)\sin({\angle {X_2 O X_1} })} \\
  B &=& \sqrt{\frac{1}{4}(X_1^2+X_2^2)\mp \frac{1}{2}(X_1X_2)\sin({\angle {X_2 O X_1}})}.
\end{eqnarray}

Now, the problem is to choose which amplitude is for Alice. Since the probability that the values
for A and B are the same is extremely low, Alice can obtain her amplitude if she knows whether her
amplitude is greater than Bob's one. By a known pilot bit sequence from the interference-free
part, which is added at the beginning and the end of each packet to distinguish whose packet it
is, Alice can align her signal with received signal through matching a known bit sequence. As a
result, she knows what is the next bit for her packet.

The transformation of structure in interfered samples occurs when the next bits for Alice and Bob
are different. We use that information to choose the right amplitude. Mainly there are two cases.

(1) The next bit for Alice signal is 1. If the amplitude of Alice's signal is greater than that of
Bob's signal, the angle difference between $X_1$ and $X_2$ varies from 0 to $\pi$; otherwise it
varies from $\pi$ to $2\pi$.

The reason is that the diagonal of a parallelogram is closer to the line which is greater than the
other. In that sense, if the two next vectors do not cross each other when they change from the
previous vectors like Fig. \ref{fig:7}(a), the range of the angle between $X_1$ and $X_2$ is from
$\frac{\pi}{2}$ to $\frac{3\pi}{2}$. If additionally the amplitude of Alice's signal is greater
than that of Bob's signal, then the angle varies from $\frac{\pi}{2}$ to $\pi$; otherwise varies
from $\pi$ to $\frac{3\pi}{2}$.

If the next vectors for Alice and Bob signals cross each other like Fig. \ref{fig:7}(b), then the
angle varies from 0 to $\frac{\pi}{2}$ degree when the amplitude of Alice's signal is greater than
Bob's one. If the amplitude of Alice signal is less than Bob's one, then the angle varies from $-
\frac{\pi}{2}$ to 0.

\begin{table}[ht]\label{Table1}
\caption{The example when the next bit for A is 1} \centering
\begin{tabular}{c c c c}
\hline \hline
Case & $A > B$ & $B > A$ \\
\hline
No crossing & $\frac{\pi}{2} \sim \pi$ & $\pi \sim \frac{3\pi}{2}$ \\
Crossing & $0 \sim \frac{\pi}{2}$ & $\frac{3\pi}{2} \sim 2\pi$ \\ \hline
\end{tabular}

\end{table}

Table I shows the examples for the four cases where crossing means that the vectors for A and B
cross each other when their phases are moved according to the next bits. Fig. \ref{fig:7} in our
paper shows that (a) is the no-crossing example and (b) is the crossing one.

(2) The next bit for Alice signal is 0. Similarly, when the amplitude of Alice's signal is greater
than that of Bob's signal, the angle between $X_1$ and $X_2$ is in range from $\pi$ to $2 \pi$;
otherwise it is from 0 to $\pi$.

In this way, we can get values for $A$ and $B$. If we take many transformation samples and average
$A$ and $B$, the averaged amplitudes will be more accurate.\newline

Now, a practical issue is how Alice detects the transformation of the interfered signal. When the
transformation occurs, the energy of received signal changes from $\mu + \sigma$ to $\mu -
\sigma$, or vice versa, where $\sigma$ is measured one (\ref{eq:sigma}). Namely, the energy
difference is $2\sigma=4AB|\cos(R)|$. Our detecting algorithm declares occurrence of the
transformation of the structure when the energy difference between the previous and the current
signals $(X_1, X_2)$ is greater than $(\sigma - \mu)=2AB|\cos(R)|$. Since there is some noise, the
variance does not change exactly $\pm(\sigma - \mu)$. Therefore, $(\sigma - \mu)$ is a relevant
threshold but the other threshold can be applied with regard to the SNR.

\section{Simulation Results}

We have simulated it using Matlab which is commonly used for simulation in variety areas. Although
Matlab is not enough to obtain a correct benefit of ANC, it can at least show whether the new
approach to estimate amplitudes from the interfered signal works properly.

In this simulation, we have assumed some points. First, the interference detector described in
\cite{1} determines the interference starting point without error. Second, the first bit of 64bit
length known pilot sequence, which is added for practical issue \cite{1}, is a fixed bit. Since
this models flat-fading quasi-static channels, there is high probability to make a wrong decision
for the first bit. Last, in \cite{1}, they have already provided that the average overlap between
Alice's packet and Bob's packet is 80\%. We just use it because it is difficult to set the size of
the time slot used for random delay. In a similar sense, we have not added the header, which
includes destination information.

We have simulated ANC with the new method in the SNR region of $20-30$ dB, which is a typical
range for 802.11 \cite{13,14}. We analyse the relationship between SNR, signal-to-interference
ratio (SIR) and bit error rate (BER) at the same time for ANC. The simulation was implemented in
the SNR range of from 20dB to 30dB and the SIR range of from -3dB to 3dB. The SIR is defined as,
\begin{center}
    $\text{SIR}=10\log_{10}{\frac{P_{Alice}}{P_{Bob}} }$
\end{center}

\begin{figure}
\centering
\includegraphics[scale=0.4]{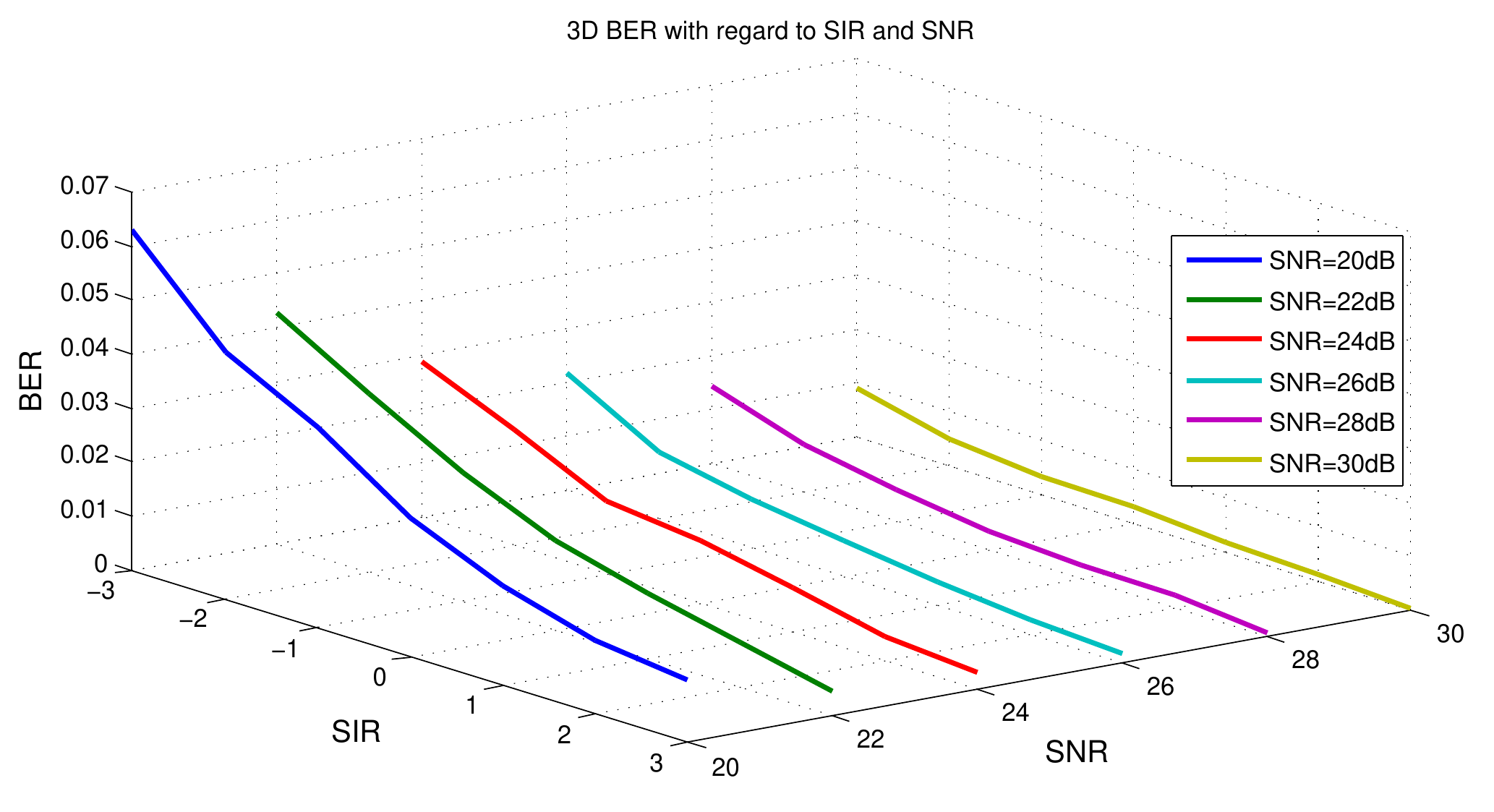}  \caption{3D BER graph with respect to SNR and SIR.}\label{fig:13}
\end{figure}

This result in Fig. \ref{fig:13} shows that ANC works well even though two transmission powers are
different. However, the best approach would be to use the equal transmission power for Alice and
Bob's signals. The reason is that when the SIR is 3dB so that Alice can decode received signal
with the BER of around 1\% which is easy to correct, Bob will face the BER of approximate 6\%
which is relatively high.

The result is not much different from the original result in \cite{1}. The original result has
provided that the average BER is approximate 4\% and the BER varies from around 4.5\% to below 1\%
dependent on SIR from -3dB to 3dB.

\section{Concluding remarks}

We have proposed a new noncoherent decoder for ANC which jointly estimates the two amplitudes from
interfered MSK signals. It is of interest to extend the idea to other modulation formats and
network topologies.

\bibliographystyle{IEEEbib}%
\bibliography{thesis}

\end{document}